\documentclass[10pt]{article}
\usepackage[top=0.85in,left=2.75in,footskip=0.75in]{geometry}

\usepackage{changepage}

\usepackage[utf8]{inputenc}

\usepackage{textcomp,marvosym}

\usepackage{fixltx2e}

\usepackage{amsmath,amssymb}

\usepackage{cite}

\usepackage{nameref,hyperref}

\usepackage[right]{lineno}

\usepackage{microtype}
\DisableLigatures[f]{encoding = *, family = * }

\usepackage{rotating}

\raggedright
\usepackage[aboveskip=1pt,labelfont=bf,labelsep=period,justification=raggedright,singlelinecheck=off]{caption}

\makeatletter
\renewcommand{\@biblabel}[1]{\quad#1.}
\makeatother

\date{}

\usepackage{lastpage,fancyhdr,graphicx}
\usepackage{epstopdf}
\fancyheadoffset[L]{2.25in}
\fancyfootoffset[L]{2.25in}
\newcommand{\be}{\begin{eqnarray}}
\newcommand{\ee}{\end{eqnarray}}
\newcommand{\ba}{\begin{array}}
\newcommand{\ea}{\end{array}}
\newcommand{\bv}{\begin{verbatim}}
\newcommand{\ev}{\end{verbatim}}
\newcommand{\bt}{\begin{tabular}}
\newcommand{\et}{\end{tabular}}
\newcommand{\btab}{\begin{table}}
\newcommand{\etab}{\end{table}}
\newcommand{\bfig}{\begin{figure}}
\newcommand{\efig}{\end{figure}}
\newcommand{\bc}{\begin{center}}
\newcommand{\ec}{\end{center}}
\newcommand{\bit}{\begin{itemize}}
\newcommand{\eit}{\end{itemize}}
\newcommand{\bi}{\begin{itemize}}
\newcommand{\ei}{\end{itemize}}

\newcommand{\Sum}{\displaystyle\sum\limits}
\newcommand{\tw}{\textwidth}
\newcommand{\ig}[1]{\includegraphics[width=#1\tw]}

\begin{document}

\vspace*{0.35in}

\begin{flushleft}
{\Large
\textbf\newline{Large scale evaluation of differences between network-based and pairwise sequence-alignment-based methods of dendrogram reconstruction}
}


Daniel Gamermann\textsuperscript{1},
Arnau Montagud\textsuperscript{2},
J. Alberto Conejero\textsuperscript{3},
Pedro Fern\'andez~de C\'ordoba\textsuperscript{3},
Javier F.~Urchuegu\'{\i}a\textsuperscript{4}

\bigskip
\bf{1} Instituto de Física, Universidade Federal do Rio Grande do Sul (UFRGS), Av. Bento Gonçalves 9500, CP 15051, 91501-970 Porto Alegre RS, Brazil
\\
\bf{2} Institut Curie, PSL Research University, Mines Paris Tech, INSERM, U900, 26 rue d'Ulm, F-75005, Paris, France
\\
\bf{3} Instituto Universitario de Matem\'atica Pura y Aplicada - IUMPA, Universidad Polit\'ecnica de Valencia, E-46022 Valencia, Spain
\\
\bf{4} Instituto Universitario de Tecnolog\'ias de la Informaci\'on y Comunicaciones - ITACA, Universidad Polit\'ecnica de Valencia, E-46022 Valencia, Spain
\bigskip

* aconejero@upv.es

\end{flushleft}
\section*{Abstract}
Dendrograms are a way to represent evolutionary relationships between organisms. Nowadays, these are inferred based on the comparison of genes or protein sequences by taking into account their differences and similarities. 
The genetic material of choice for the sequence alignments (all the genes or sets of genes) results in distinct inferred dendrograms. 
In this work, we evaluate differences between dendrograms reconstructed with different methodologies and obtained for different sets of organisms chosen at random from a much larger set. A statistical analysis is performed in order to estimate the fluctuation between the results obtained from the different methodologies. 
This analysis permit us to validate a systematic approach, based on the comparison of the organisms' metabolic networks for inferring dendrograms. It has the advantage that it allows the comparison of organisms very far away in the evolutionary tree even if they have no known ortholog gene in common.


\section*{Introduction}

Dendrograms are a way to represent the evolutionary relationships among entities, such as species, proteins, coding genes, exons,...
In our case, for a given dendrogram we will consider two types of nodes: Leaves (a node connected with another single node) represent species, either current or extincted, and the rest of nodes (connected with more than one node) represent a common ancestor of the nodes hanging from it. 

These dendrograms can only be inferred based on data of currently living species or, in a few cases, using fossil records. Currently, the most accepted methodology in order to construct (infer) such dendrograms is to infer the distance of two organisms to their common ancestor based on the comparison (alignment and scoring) of their genetic sequences, what is termed phylogenetics.

Alignments between sequences are not unique, as the scoring of the alignments can differ. As a consequence, different dendrograms will be reconstructed for the same set of organisms when applying different methodologies (e.g. distance matrix, maximum parsimony, maximum likelihood, Bayesian inference,\ldots) in the reconstruction. Even the same methodology may result in different dendrograms depending on the material one chooses to study, e.g. a single gene, a set of genes, amino acid sequences or whole genomes. 
Therefore, it may become important not only to obtain a dendrogram, but to know how a dendrogram can be compared to others. 
In other words, a measurement to compare several dendrograms and to evaluate what is the expected difference between them and their fluctuation is needed. 
An accepted metric in order to evaluate dendrograms' differences is the Robinson-Foulds metric \cite{robinson1981} also known as the symmetric difference metric on dendrograms, which evaluates the cost needed to modify one dendrogram to obtain the other. See also \cite{day1985optimal,pattengale2007,bocker2013}.

Closely related species share many genes in common, while distant species will share very few traits. Traditionally, phylogenetic relationships among distant species have been computed using the small subunit ribosomal RNA (16S) sequences in the comparisons \cite{woese1977}. 
Some works have used other conserved sequences, such as a subset of genes \cite{ciccarelli2006} or a combination of these \cite{lienau2010}. In the last years it has been increasingly feasible to perform whole genome alignments \cite{wu2008, wu2009}.
Studies have pointed out the importance of considering only sets of genes \cite{rokas2003}, but they have been mixed about the usefulness of filtering the genome sequences that are compared \cite{jeffroy2006}. 
Thus, which is the perfect set of sequences, if any, in order to obtain a dendrogram that includes very distant species is still a matter of debate \cite{lienau2010}.

Recently, a new approach based on the comparison of metabolic networks has been proposed in order to infer the distance between two organisms \cite{gamermann2014}. 
Metabolic networks are graphs where every metabolite in an organism's metabolome represents a node and pairs of nodes are connected whenever a chemical reaction in the organism's metabolism connects the two metabolites as substrate-product. Metabolic networks' properties have been extensively studied \cite{jeong2000} and present many characteristics in common (e.g. approximate scale-free distribution of their node's degrees, high clustering coefficient, small-world structure), which indicate a common internal organization of the studied metabolisms.

A metabolic network is reconstructed using the information of all enzymes contained in an organism. Therefore, it contains the information of a large subset of this organism's genome. Moreover, even organisms far away in the evolutionary tree will share important pathways; also, many metabolites (nodes) are ubiquitous and will be present in all species. This explains that differences and similarities can always be established between two given metabolisms. In fact, is has been published that the comparison of metabolic networks represents a valuable tool in order to infer phylogenetic relationships \cite{clemente,gamermann2014,deyasi2015}.

In this work, we systematically construct and compare dendrograms built from different sets of organisms using different genes, proteins or networks. Our goal is to present evidences that dendrograms reconstructed using only information from metabolic networks are comparable to more traditional gene-based dendrograms in terms of accuracy and comprehensiveness. 

The work is organized as follows: In the \textit{Materials and Methods} section, we explain in detail how we obtained and processed our data in order to reconstruct the dendrograms, how the sequence alignments were performed and the scoring systems and methods we used to obtain the distance matrices and, lastly, how to evaluate the distances among dendrograms. We also explain the graph theoretical aspects used in the network comparison, how the ``network'' dendrograms were constructed and how the dendrograms' differences were evaluated. In the \textit{Results and Discussion} section we explain the statistical analysis performed and discuss our results. We also included an appendix with mathematical details about 
how the Pagerank algorithm is used to determine the relative importance of every metabolite in an organism based on their connections to the rest of the metabolic network.

\section*{Materials and Methods}
\subsection*{Dataset used to build the dendrograms}

We retrieved from the KEGG database \cite{kegg} a large set of organisms genes, and we identified those associated with enzymes. For each enzyme in a given organism, we identified all the chemical reactions associated with that enzyme, such that, for each organism we were able to build a list of all identified chemical reactions potentially present in its metabolism. Moreover, for each gene we obtained their corresponding nucleic acid and amino acid sequences. Details on the procedures used to obtain information from KEGG can be found in \cite{raymari}.

Separately, for each prokaryotic organism in our dataset, we searched the NCBI database for its 16S rRNA subunit sequence using an automatized script including the terms \texttt{Genus species[Orgn] AND 16S ribosomal RNA[Titl] NOT partial sequence[Titl]}, where \texttt{Genus species} was the binomial nomenclature of each organism in the dataset obtained from KEGG. This search was designed to retrieve only complete sequences and discard partial ones. 

Our original data set built with KEGG's information comprised originally 4803 organisms. From these, the metabolic networks of 3972 organisms were completed, whereby NCBI searches retrieved 16S rRNA subunit sequences for 1537 of them. The intersection of all these sets resulted in a dataset with 1506 prokaryote organisms for which we had complete information, i.e. we had all sequences for their enzymes, the complete list of chemical reactions and 16S rRNA nucleotide sequences.

\subsection*{Definition and construction of dendrograms}

Our analysis is based on three categories of dendrograms. Shortly, these are referred to as \textit{gene-based dendrograms}, \textit{network dendrograms}, and \textit{random dendrograms}. Gene-based dendrograms are those constructed based on sequence alignments. We compute three different gene-based dendrograms, the difference between them comes from the sequence (or sequences) used in the alignments: either a large set of proteins (amino acid sequences); a single protein from this set; or the 16S rRNA subunit nucleic-acid sequences. Metabolic network dendrograms are those constructed via comparison of metabolic networks reconstructed from the list of chemical reactions that is obtained from the annotation of the organism's genome. The random dendrograms are constructed by linking the organisms in a set at random.

Given a set of $N$ organisms the first step in our proposed dendrogram reconstruction is the evaluation of a symmetric $N\times N$ distance matrix ($D$), where each element element $D_{ij}$ is a measure of the ``evolutionary'' distance between organism $i$ and $j$. The evaluation of this matrix follows different methodologies that are described in the following subsections. Here we explain the reconstruction of the dendrogram once the $D$ matrix is calculated. For this, we follow the same procedure as in \cite{gamermann2014}.

The matrix $D$ can be viewed as a complete weighted graph $G=(V,E,w)$. The set of nodes $V$ stands for all the organisms in the dataset. Each pair of different organisms are linked by an edge in $E$. 
A non-negative function $w:E\rightarrow\mathbb{R}_0^+$ associates a weight to each edge, according to the distance between the organisms connected by that edge. Once this weighted graph is generated, we apply Kruskal algorithm to obtain a minimum spanning tree. A \textit{spanning tree} is an acyclic and connected subgraph $G'=(V'E',w')$ of $G$ such that $V'=V$ and $E'\subset E$. The edges in $E'$ have the same weights as the corresponding ones in $E$. Among all the spanning trees of a given graph $G$, a \textit{minimum spanning tree} is a spanning tree such that the sum of the weights associated to their edges is minimum respect to all the admissible spanning trees of $G$. Further information on trees and graphs can be found in \cite{gross_yellen}. From the minimum spanning tree a \textit{dendrogram} is obtained. This dendrogram represents the relationships among the given set of $N$ organisms. The lengths of the branches in the \textit{dendrogram} are proportional to the distances in the matrix $D$. 

\subsection*{Gene-based dendrogram construction}

Gene-based dendrograms will be based on pairwise alignment of nucleotides or amino acid sequences, i.e. the matrix distance $D$ for the organisms present in a set is evaluated from the result obtained from sequence alignments. The alignments are done using the \textit{Needleman-Wunsch} algorithm \cite{needleman1970} with affine gap penalty. The algorithm inserts gaps in the sequences in order to create the alignment that maximizes some score $S$. In the scoring of an alignment the opening of a gap subtracts 10 points from $S$ and every extension of the gap subtracts 0,5 points. In the nucleotide alignments every match of nucleotides adds 5 points and a mismatch subtracts 4 points, while for the alignment of amino acid sequences, different standard matrices are used (BLOSUM and PAM). Given the alignment score $S$ we define the parameter $P$ as:

\be
P &=& 1-\frac{S}{M} \label{eq:PPP}
\ee
were $M$ is the maximum score possible (the score which would be obtained with no mismatches and no gaps in the alignment). The smaller $P$ is, the closer the two sequences are. Typically, $P$ is a value between 0 and 1 but, for very bad alignments, a $P$ larger than 1 is possible, meaning that gaps and mismatches in the alignment subtracted more points than matches added.

In the comparison of two organisms $1\le i,j\le N$, if each organism has only one sequence to be compared, the distance $D_{ij}$ between both of them is just the result for $P$ in (\ref{eq:PPP}) obtained from the alignment of this sequence. If one or both organisms in a comparison have more than one sequence corresponding to the same gene we match each sequence from the organism with the least number of sequences to its best alignment with sequences from the other organism. Then, we set the distance $D_{ij}$ as the average $\bar{P}$ for the values of $P$ obtained from each possible alignment. 

Three different gene-based dendrograms were constructed for each set of organisms, we call them by \textit{DRIBO}, \textit{DENZS} and \textit{D1ENZ}:

\bi
\item \textit{DRIBO} is a dendrogram constructed using the rRNA sequences for the 16S ribosomal subunit.
\item \textit{DENZS} is a dendrogram constructed using the amino acid sequences of all proteins associated to all EC numbers common to all organisms in a set. (The average number of common EC numbers among all organisms in a set, for the organisms sets we worked with, was $40.15\pm20.73$.)
\item \textit{D1ENZ} is a dendrogram constructed using the amino acid sequences associated to a single EC number taken at random from all EC numbers common to all the organisms in the set.
\ei

\subsection*{Network dendrogram construction}

For the construction of dendrograms based on networks, the matrix distance $D$ is obtained from the comparison of the metabolic networks of each pair of organisms in the set. The first proposed methodology to construct dendrograms based on metabolic networks can be found in \cite{gamermann2014}. In this previous work, a parameter ($\zeta$) is defined as the result of the comparison of two networks. This parameter depends on weighted averages over different sets of metabolites (common or not to each pair of organisms), where the weights of the metabolites are evaluated according to their connectivity degree. In the present work, we will test an array of parameters, including this $\zeta$ , in order to establish the one that produces dendrograms that are more similar to the ones produced by the other methodologies. 

Given two arbitrary organisms $1\le i,j\le N$, we consider the metabolic networks of each one of these organisms as weighted graphs. In any of these graphs, nodes stand for metabolites and edges between a pair of nodes indicate the presence of a chemical reaction in the corresponding organism's metabolism, that links the two metabolites as substrate-product. 

A successful approach to measure the importance of a node in a network can be obtained by using the Pagerank algorithm \cite{brin1998}. This was inspired by the eigenvalue problem on scientometrics and successfully used in the former versions of the Google browser. Afterwards, Pagerank has been extensively used in network theory for different purposes. For instance, in computational biology it has been used for determining which are the key species in a food web that can cause the collapse of the entire system \cite{allesina2009} or for improving outcome prediction for cancer patients \cite{winter2012}. In our work, Pagerank is used to assign weights to the edges as a result of an application of this algorithm to the graph resulting of the union of the metabolic networks of all organisms in a set. For details about our implementation of the Pagerank algorithm, please refer to the supplementary material section.

From the metabolic networks of organisms $i$ and $j$, let us define the sets of edges $A_{ij}$, $B_{ij}$ and $C_{ij}$, where $A_{ij}$ is the set of edges present in organism $i$ but not in $j$, $B_{ij}$ is the set of edges present in organism $j$ but not in $i$, and $C_{ij}$ is the set of edges present simultaneously in both networks of organisms $i$ and $j$. 

Given these three sets, $A_{ij},B_{ij},$ and $C_{ij}$ let us define the following parameters:

\be
\alpha_{ij} &=& \Sum_{l\subset A_{ij}} w_l \label{eq:alpha}\\
\beta_{ij} &=& \Sum_{l\subset B_{ij}} w_l \label{eq:beta}\\
\gamma_{ij} &=& \Sum_{l\subset C_{ij}} w_l \label{eq:gamma}
\ee
where the sums are made for the weights $w_l$, given by the Pagerank, of all edges $l$ in each set. Details in how the weights are evaluated are discussed in supplementary materials. Defined as such, the parameters $\alpha_{ij}$ and $\beta_{ij}$ represent measures of the differences between the networks $i$ and $j$ respect to each other, while the parameter $\gamma_{ij}$ is a measurement of the similarity between them. For a schematic representation of this, please refer to Figure \ref{fig:esquema1}.

Different network dendrograms are going to be constructed for each set of organisms, based on different choices of parameters for the distance matrix $D$:

\bi
\item \textit{DS1} is obtained when the distance matrix is given by $D_{ij} = |n_i-n_j|$ where $n_i$ is the number of nodes in each network.
\item \textit{DS2} is obtained if $D_{ij} = |e_i-e_j|$, where $e_i$ is the number of links in each network.
\item \textit{DNET1} is obtained if $D_{ij} = \frac{n_{tot}}{\gamma_{ij}}$, where $n_{tot}$ is the number of common metabolites in networks $i$ and $j$.
\item \textit{DNET2} is obtained if $D_{ij} = \alpha_{ij}+\beta_{ij}$, where $\alpha_{ij}$ and $\beta_{ij}$ are defined in \eqref{eq:alpha} and \eqref{eq:beta}.
\item \textit{DNET3} is obtained if $D_{ij} = \frac{\alpha_{ij}+\beta_{ij}}{\gamma_{ij}}$, where $\alpha_{ij},\beta_{ij}$ and $\gamma_{ij}$ are defined in \eqref{eq:alpha}-\eqref{eq:gamma}.
\item \textit{DNET4} is obtained if $D_{ij} = \zeta_{ij}$, with $\zeta_{ij}$ calculated following the procedure in \cite{gamermann2014}. In this article, the parameters $\alpha$, $\beta$ and $\gamma$ are evaluated following the same principles as in this present work, but the sums in \eqref{eq:alpha}-\eqref{eq:gamma} are made over nodes and not over links and the weights of the nodes are related to their connectivity. Finally, the parameter $\zeta_{ij}$ is the equivalent to the parameter used in \textit{DNET3} above, but using nodes and not links in the evaluation. 
\ei

Note that \textit{DS1} and \textit{DS2} are two different ways of comparing the difference in size of two given networks, while the other dendrograms take into account a measurement of the importance of the links and/or nodes which are either common to both networks or particular to only one of them. Additionally to these dendrograms, we will also consider a dendrogram built linking the different species at random which we will call DRAND. These random dendrograms are produced by generating a symmetrical distance matrix whose elements are uniformly distributed random numbers.

\subsection*{Dendrogram comparisons}

Since different methods have been proposed for generating dendrograms from the same set of organisms, a measure is needed for comparing them. Robinson-Foulds metric, introduced in \cite{robinson1981}, allows to measure how similar two dendrograms are. This metric has been widely used since it is not limited to binary trees and is based on counting elementary operations which transform one dendrogram into another. The lower the difference between two dendrograms is, the more similar the two dendrograms are. A more detailed description can be found in the Supplementary Material. Several algorithms have been described for efficiently compute this metric \cite{day1985optimal, pattengale2007}. In this work we have considered the implementation in the Python library DendroPy \cite{sukumaran2010}.

Two ensembles were constructed by randomly choosing organisms from the 1506 organisms set for which there was complete information. The first ensemble contains 10 sets of organisms, each set containing 20 organisms. The second ensemble contains 10 sets of 30 organisms. Additional file \texttt{ensembles.txt} contains the organisms in each set in each ensemble. In the additional files, each organism is identified by its KEGG code (usually a 3 letter code).

The procedure adopted is the following: given an ensemble, for each organisms set in the ensemble, the different distance matrices are calculated and gene-based and network dendrograms are constructed. So, for each set, 9 distance matrices (3 gene-based and 6 based on networks) are evaluated and the corresponding 9 dendrograms are constructed. Each dendrogram is compared  using the Robinson-Foulds metric to the other dendrograms, making up to 36 comparisons (as there are 36 possible combinations of 9 elements two by two). This is repeated for each set in the ensemble and the resulting comparisons are averaged over all sets.

Note that the distance parameter in each methodology has arbitrary units. For comparing the dendrograms, we rescale the distances in the dendrograms such that the biggest distance is always 1. Note that distances do not have a direct correspondence to any real unit, only the relative distance has a meaning. Therefore, a rescaling of the numbers in a dendrogram should not result in any bias in the comparisons.

Figure \ref{fig:workflow} illustrates the workflow adopted: we have picked at random the sets of organisms to build up both ensembles, then we have compared their sequences and with Kruskal algorithm we have built dendrograms. Then we have compared the different dendrograms using Robinson-Foulds metric.

\section*{Results and Discussion}

We have worked with two ensembles, all constructed by randomly selecting organisms from the 1506 organisms dataset for which we had complete information (see the aforementioned dataset subsection). The first ensemble contained ten organisms sets with twenty organisms in each set while the second ensemble contained ten organisms sets with thirty organisms in each set.

For each set in an ensemble, we construct 3 gene-based dendrograms (denoted by DRIBO, DENZS and D1ENZ), 6 network dendrograms (denoted by DS1, DS2, DNET1, DNET2, DNET3 and DNET4) and also 100 random dendrograms (DRAND). Then we compare each dendrogram with each random dendrogram and with each other (calculate the symmetric difference, i.e. Robinson-Foulds Metric). We evaluate the average and standard deviation for every pair of PT for each organisms set, so that all comparisons are covered.

The results of these averages are in Tables \ref{tab1}-\ref{tab3} with the standard deviation given as uncertainty. The smaller the value in an element in one of these tables is, the more similar the corresponding dendrograms are. In an additional file \texttt{trees.txt}, we provide all 9 dendrograms (DRIBO, DENZS, D1ENZ, DS1, DS2, DNET1, DNET2, DNET3, DNET4) obtained for each set in each ensemble.

We wanted to compare our dendrograms built with enzymatic information with well known distances of amino acid substitutions. Thus, we compared the DENZS, a dendrogram constructed using all common enzymes amino acids sequences, built using different scoring matrices, such as BLOSUM and PAM matrices. BLOSUM matrices are amino acids substitution matrices based on observed alignments \cite{henikoff1992}. BLOSUM45 is used for distantly-related proteins and BLOSUM62 for midrange-related proteins. On the other hand, PAM amino acids substitution matrices' observations are extrapolated from comparisons of closely related proteins, as they look for point accepted mutations (PAM) \cite{dayhoff1978}. These consist on the replacement of a single amino acid in the protein sequence with another single amino acid. For instance, PAM250 matrix was calculated based on 1572 observed mutations in 71 families of proteins with alignments that were more than 85\% identical \cite{PAMmatrix}. Results can be seen in Table \ref{tab1}. Unsurprisingly, this table shows small distances between DENZS dendrograms built with different substitution matrices and, thus, the resulting dendrograms are very similar. This is due to the fact that PAM and BLOSUM matrices have equivalences, for instance, PAM250 retrieves very similar results as BLOSUM45 \cite{PAMmatrix} and, thus, dendrograms built with substitution matrices that give similar results will be similar. From this diversity of DENZS dendrograms, we chose to use for the following comparison only the DENZS built with the BLOSUM55 matrix.

In order to visualize the comparison of results, dendrograms were done from the tables using \textit{pvclust} package \cite{Suzuki2006} in R statistical software using Ward.D clustering method and euclidean distance on the Robinson-Foulds values for each dendrogram pair. Two different methods of significance are shown: approximately unbiased p-value (AU) and bootstrap probability value (BP). AU p-value is computed by multiscale bootstrap resampling and is generally a better approximation to unbiased p-value than BP value that is computed by normal bootstrap resampling \cite{Suzuki2006}.

In Table \ref{tab2} and Figure \ref{fig:dendo1} we have compared gene-based and network methods for the first ensemble with ten sets of twenty organisms each and in Table \ref{tab3} and Figure \ref{fig:dendo2} we have shown the same results for the second ensemble with ten sets of thirty organisms each. As expected, values in these tables are higher than in table \ref{tab1} where the only difference in the construction of the dendrograms is the scoring matrices used in the alignments. Due to its usage as the closest measure we have to an evolutionary distance standard regarding pair-wise sequence comparison, we have compared our results to DRIBO dendrograms, which has been constructed comparing 16s ribosomal sequences among organisms within each set. 
In the first ensemble, DNET1 and DNET2 are the closest dendrograms to DRIBO, very close to the following two: DNET2 and DNET3. In the second ensemble, DNET2 is closest to DRIBO, but very close to DENZS and D1ENZ. In both cases dendrograms built using network (DNET1, DNET2, DNET3 and DNET4) and enzymes (D1ENZ and DENZS) information are closer to DRIBO than dendrograms built using number of nodes (DS1) or links (DS2) information or randomly built (DRAND).

\section*{Conclusions}

Building dendrograms is an approximation to capture the evolutionary distance of different species. Present work targets the potential of using metabolic topologies of two given species to search for evolutionary distances rather than (or as a complementary way) to use pair-wise sequence comparison of enzymes. The results of the second ensemble suggest that, in some cases, network comparison might be even better than amino acid sequence alignment of enzymes in order to infer relationships between organisms. Also of importance is the message that considering the size of the networks as a distance between organisms is a very poor way to capture the organisms' relationship, as can be seen with the results for dendrograms DS1 (number of nodes in the network) and DS2 (number of links), that are closer to DRAND than to gene-based dendrograms.

The last decade has provided researchers with loads of sequences from a wide variety of organisms, promoting the development of new tools and the renewal of old ones. Hereby, we have have shown the possibility to incorporate topological information in these studies, as well as to compare dendrograms built with very different methodologies and to study their ability to capture evolutionary distances comparing it with a well established methodology, such as the alignment of the 16S subunit of ribosomal RNA. This shows the potential of network studies to explain and complement sequence alignment methodologies and hints to a future where evolutionary distances may be calculated considering very different sources of information. This complementarity has been the focus of a recent work where metabolic networks and evolution have been shown to give very interesting insights into one another \cite{naturearn}.

\section*{Acknowledgments}

The research leading to these results has received funding from the European Union Seventh Framework Program (FP7/2007-2013) under grant agreement number 308518 (CyanoFactory).

We also thanks Salvador Capella-Guti\'errez for helpful discussions on the topic.

\section*{Authors' contribution}
D.~Gamermann and J.A.~Conejero developed the methodology, performed the calculations and built the dendrograms. A.~Montagud analysed the results and these three authors wrote the manuscript. P.~Fernández de C\'ordoba and J.~Urchuegu\'ia conceived and funded the study. All authors read and approved the manuscript.


\bibliography{trees}


\newpage
\section*{Supplementary Material: The Pagerank algorithm}

The Pagerank algorithm is based on the existence of a unique eigenvector associated to the eigenvalue 1 for any arbitrary regular Markov transition matrix. In this case, we also have that all other eigenvalues have modulus strictly smaller than 1. The idea of Pagerank algorithm is to transform the adjacency matrix of a graph into a regular Markov transition matrix and then, to find the Perron-Fr\"obenius vector, see \cite{brin1998} and also \cite{langville2005,langville2006}.

First, given a set of organisms, we consider a new graph were the nodes correspond to metabolites present in at least one organism from the set. The set of edges represent if a pair of metabolites is connected in at least one organism. The weighted adjacency matrix $M$ indicates the frequency that each pair of metabolites are connected in different organisms. More precisely, for an arbitrary pair $i,j$, $M_{ij}$ stands for the number of times that a link connecting metabolite $i$ with metabolite $j$ appears among the organisms in the set.

Then, we define the matrix $S$ for computing the relative frequency of a connection of metabolite $i$ with metabolite $j$ respect the total number of connections of metabolite $i$ with the others metabolites in the set.:

\be
S_{ij} &=& \frac{M_{ij}}{k_j} \\
k_j &=& \Sum_i M_{ij}
\ee
To understand what matrix $S$ means, imagine a random walker in node $j$ of our network. He goes to the node $i$ with a probability given by $S_{ij}$. Let us take a column vector $\phi$, whose coordinate $i$ indicates the probability to find our walker at node $i$ of our graph at a certain  moment. All the entries of vector $\phi$ are positive and they add to 1. Then, after our walker changes its position once, the new probability vector would be given by $S\phi$. So as to, after $n$ jumps, it would be given by $S^n\phi$. If all the entries of $S$ were strictly positive, in the limit that $n$ goes to infinity we obtain a stationary probability vector $\varphi$:

\be
S\varphi &=& \varphi
\ee
which means that the vector $\varphi$ is the eigenvector of $S$ with eigenvalue equal to 1, see for instance \cite{meyer2000}. This convergence can also be deduced as a consequence of the Banach fixed point theorem \cite{langville2006}.

As we have reported, $S$ is an adjacency matrix and it usually has many entries which are null. The graph represented by the matrix $S$ may have disconnected components or bottle necks (in the case of a directed network). In these cases, the Perron-Fr\"obenius theorem that ensures the existence of a single eigenvector corresponding to the eigenvalue 1 would not stand. A trick for making the matrix $S$ strictly positive and without disconnected components or bottle necks is to introduce a probability ($1-\alpha$) of a random jump in between any two nodes of the graph by using matrix $G$, instead of $S$, defined as:

\be
G_{ij} &=& \alpha S_{ij} + \frac{1-\alpha}{n} 
\ee
where $n$ is the total number of nodes in the network and $0<\alpha<<1$.

Finally, the Pagerank of metabolite $i$ is defined as the element $i$ of the eigenvector $\varphi$ of $G$ given by:

\be
G\varphi &=& \varphi
\ee
Once computed the pagerank for all the metabolites in a set, we define the edge weights $w_l$ in Eqs (\ref{eq:alpha})-(\ref{eq:gamma}) as:

\be
w_l &=& \varphi_i+\varphi_j
\ee
where the edge $l$ connects metabolites $i$ and $j$, whose respective pageranks are $\varphi_i$ and $\varphi_j$.

In Figure \ref{fig:pgrk} we show a plot for the elements of the eigenvector $\varphi$ calculated for a set of 10 organisms chosen at random from our dataset. 

\newpage




\newpage


\section*{Figure Legends}

\bfig[!ht]
\bc
\bt{lr}
\\
\ig{0.8}{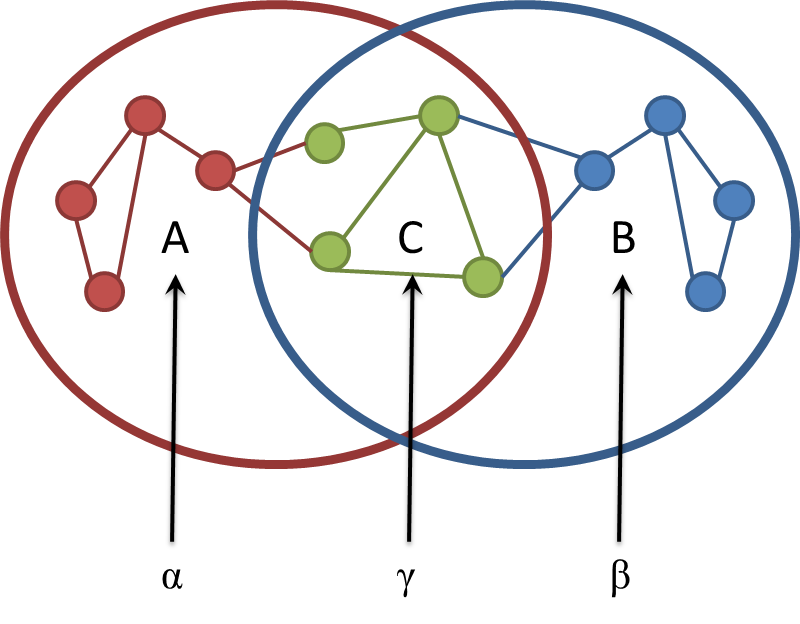} 
\et
\ec
\caption{Schematic of A, B and C sets and their parameters.} \label{fig:esquema1}
\efig

\newpage

\bfig[!ht]
\bc
\bt{lr}
Obtain dendrogram for one organism set. & Compare the difference between \\
  & dendrograms for many sets.\\
\ig{0.4}{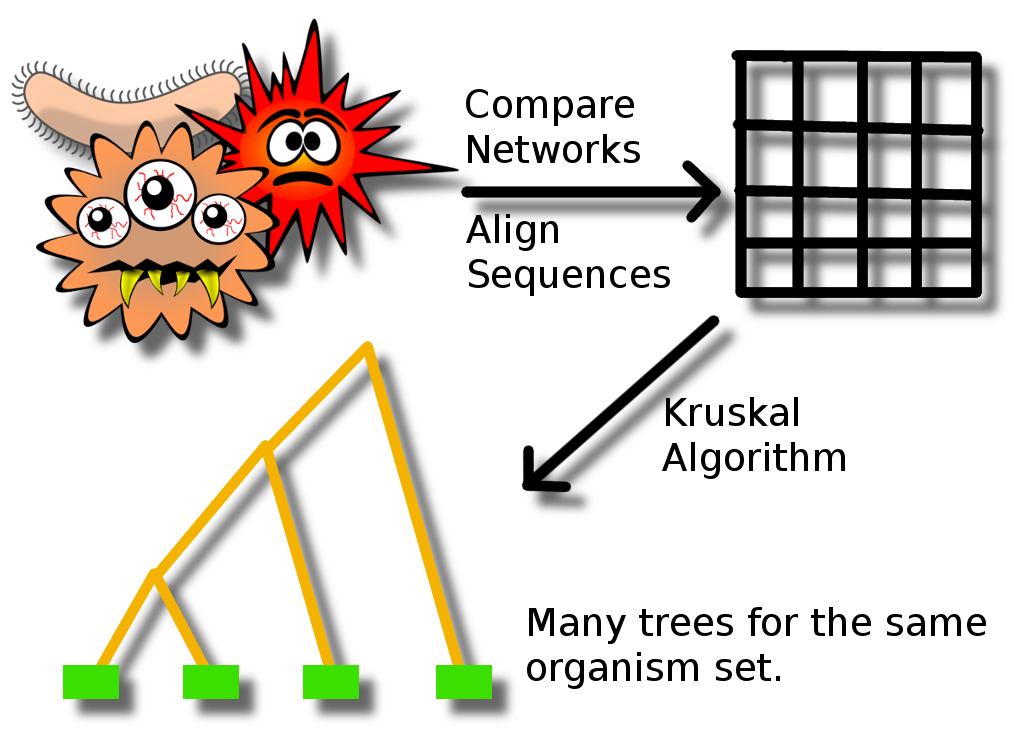} & \ig{0.4}{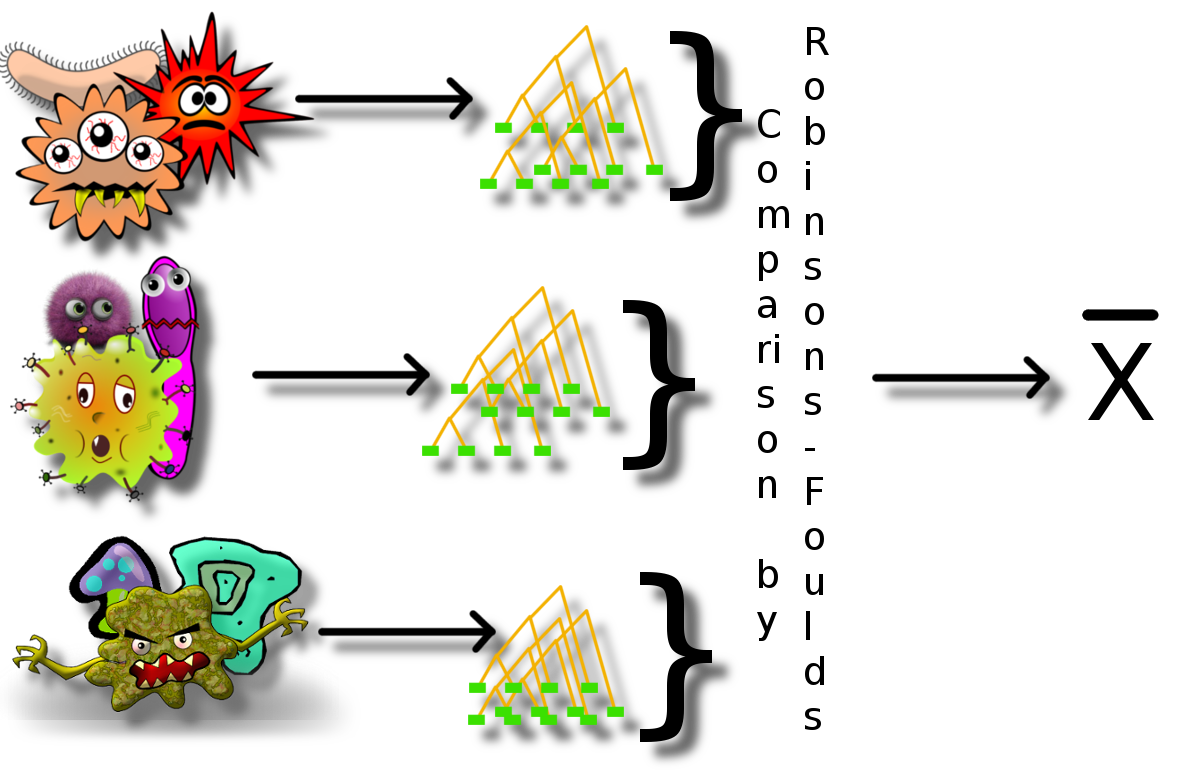}
\et
\ec
\caption{Work flow for evaluating and comparing dendrograms. (Bacteria cartoons from https://pixabay.com/)} \label{fig:workflow}
\efig

\newpage

\bfig[!ht]
\bc
\ig{0.8}{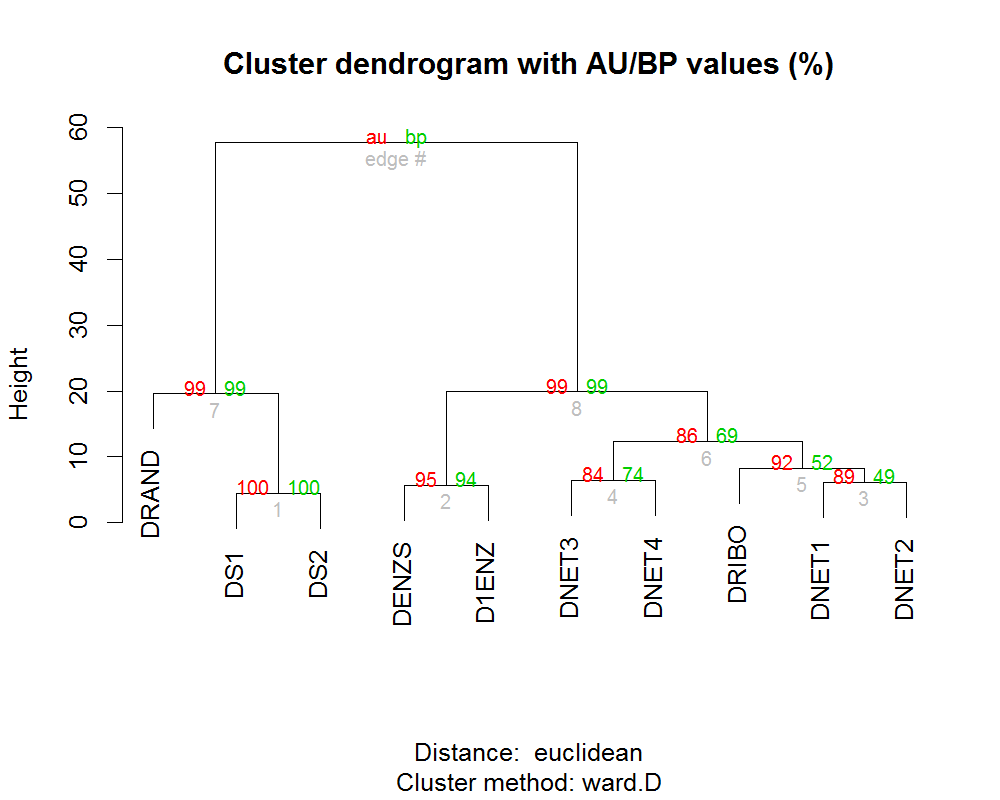} 
\ec
\caption{Cluster  of dendrograms built with different methodologies for the first ensemble of organisms.} \label{fig:dendo1}
\efig

\newpage

\bfig[!ht]
\bc
\ig{0.8}{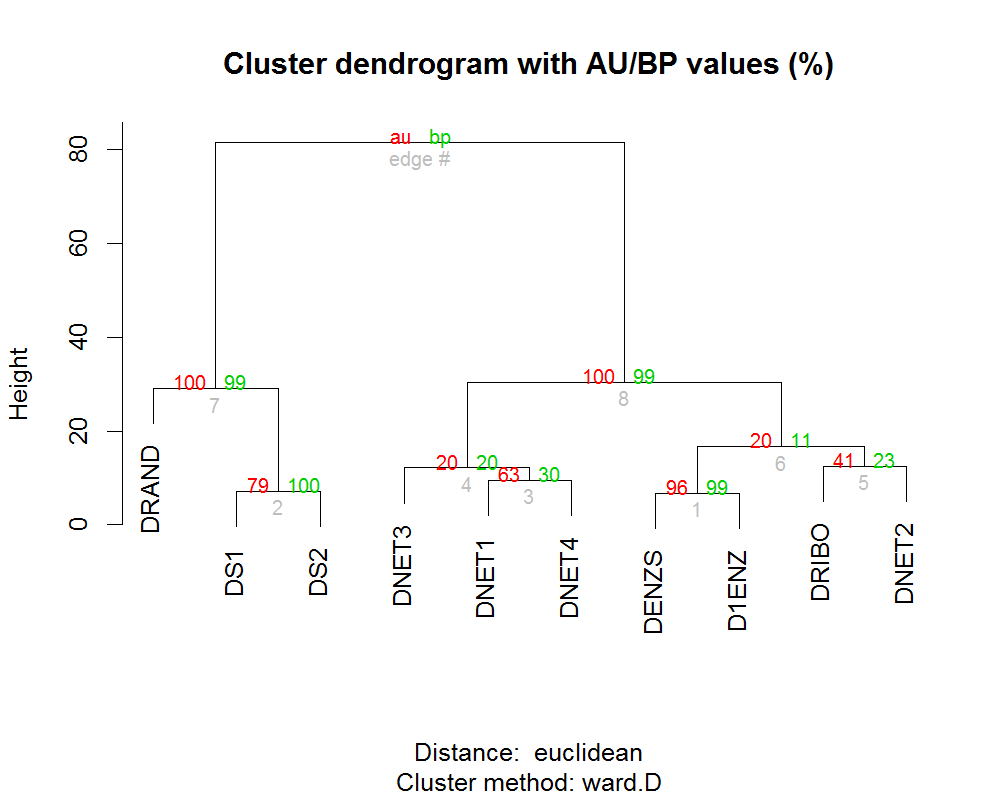}
\ec
\caption{Cluster of dendrograms built with different methodologies for the second ensemble of organisms.} \label{fig:dendo2}
\efig

\newpage

\bfig[!ht]
\bc
\bt{cc}
\ig{0.5}{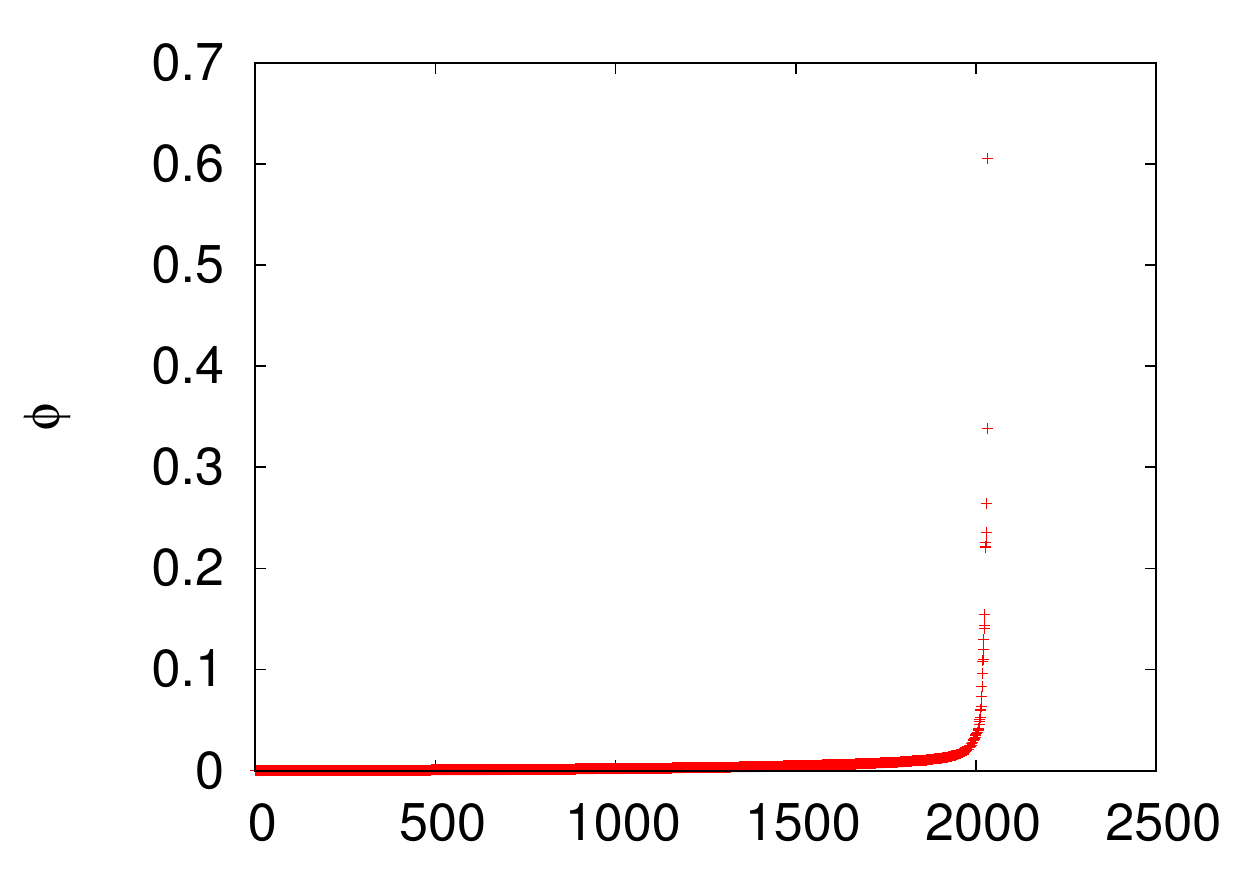} & \ig{0.5}{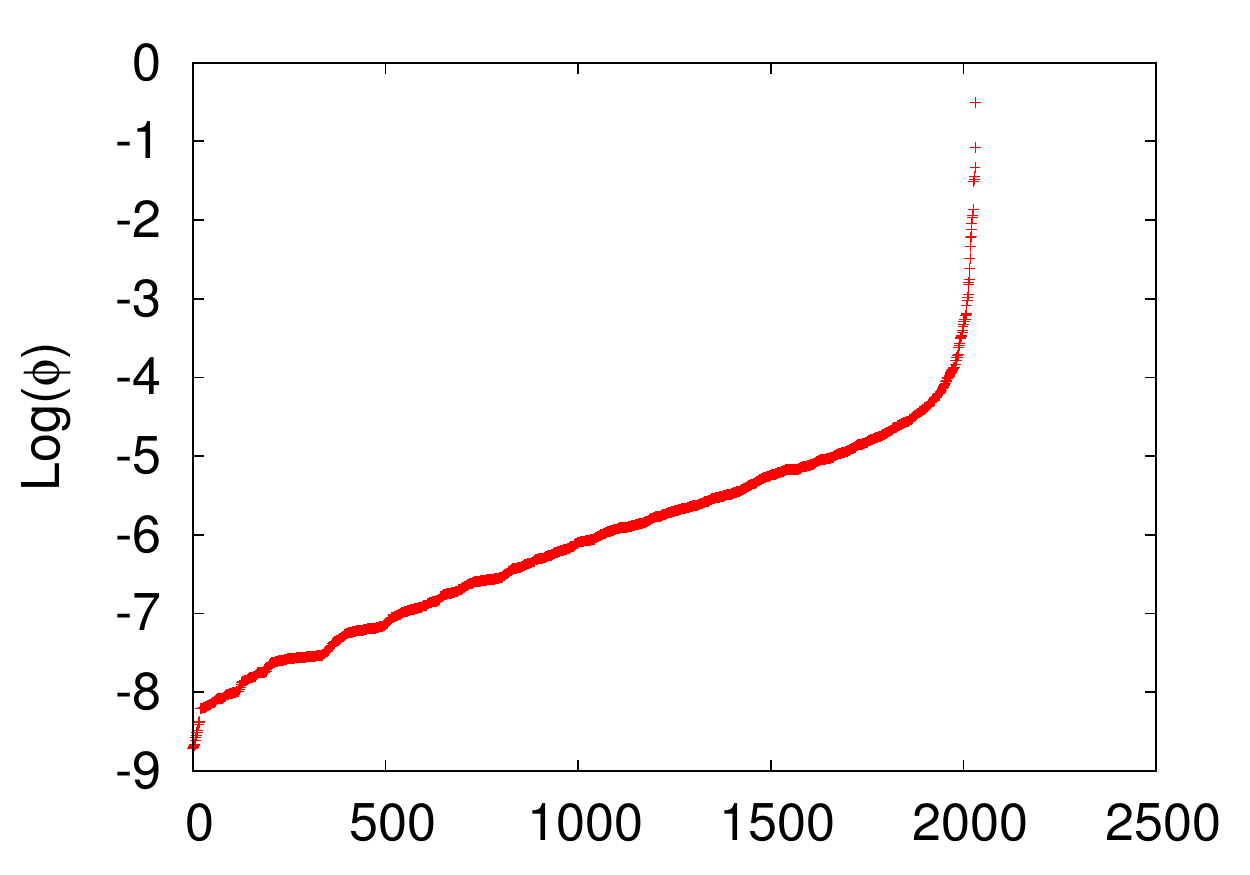}
\et
\ec
\caption{Page rank vector for $\alpha=0.99$ ordered for a set of 10 organisms chosen at random.} \label{fig:pgrk}
\efig

\newpage

\section*{Tables}
\btab[!ht]
\small
\begin{adjustwidth}{-2.25in}{0in}
\caption{\bf Comparison of DENZS dendrograms built using different scoring matrices for the first ensemble (10 sets of 20 organisms in each set).} \label{tab1}
\bt{c|ccccccc}
DENDROGRAMS &   BLO 55 &  BLO 62 &  BLO 90 &  PAM 60 &  PAM 120 &  PAM 250 &  RAND\\ 
\hline
\hline
     BLO 45  &  0.055 $\pm$ 0.025 &  0.524 $\pm$ 0.038 &  1.219 $\pm$ 0.071 &  2.082 $\pm$ 0.137 &  1.146 $\pm$ 0.104 &  0.462 $\pm$ 0.113 &  15.237 $\pm$ 0.463\\  
     BLO 55  &   &  0.516 $\pm$ 0.036 &  1.210 $\pm$ 0.068 &  2.064 $\pm$ 0.136 &  1.135 $\pm$ 0.108 &  0.480 $\pm$ 0.114 &  15.246 $\pm$ 0.462\\  
     BLO 62  &   &   &  0.721 $\pm$ 0.062 &  1.581 $\pm$ 0.123 &  0.662 $\pm$ 0.098 &  0.805 $\pm$ 0.133 &  15.733 $\pm$ 0.493\\  
     BLO 90  &   &   &   &  0.897 $\pm$ 0.080 &  0.305 $\pm$ 0.075 &  1.479 $\pm$ 0.152 &  16.383 $\pm$ 0.536\\  
     PAM 60  &   &   &   &   &  1.021 $\pm$ 0.084 &  2.344 $\pm$ 0.183 &  17.213 $\pm$ 0.596\\  
    PAM 120  &   &   &   &   &   &  1.368 $\pm$ 0.150 &  16.318 $\pm$ 0.554\\  
    PAM 250  &   &   &   &   &   &   &  15.088 $\pm$ 0.480\\  
    RAND &    &   &   &   &   &   &  \\  
\et
\end{adjustwidth}
\etab

\newpage

\btab[!ht]
\tiny
\begin{adjustwidth}{-2.25in}{0in}
\caption{\bf Comparison of Different Gene-based and Network dendrograms for the first ensemble (10 sets of 20 organisms in each set).} \label{tab2}
\bt{c|ccccccccc}
DENDROGRAMS  &  D1ENZ &  DRIBO &  DS1 &  DS2 &  DNET1 &  DNET2 &  DNET3 &  DNET4 &  DRAND\\ 
\hline
\hline
   DENZS  &  3.796 $\pm$ 1.471 &  4.631 $\pm$ 1.706 &  13.062 $\pm$ 0.402 &  13.036 $\pm$ 0.380 &  5.345 $\pm$ 1.348 &  5.612 $\pm$ 1.076 &  8.432 $\pm$ 1.345 &  7.189 $\pm$ 1.427 &  15.156 $\pm$ 0.417\\  
   D1ENZ  &   &  4.918 $\pm$ 1.964 &  12.022 $\pm$ 1.742 &  11.999 $\pm$ 1.872 &  5.687 $\pm$ 1.991 &  5.698 $\pm$ 1.642 &  7.936 $\pm$ 2.592 &  6.782 $\pm$ 1.888 &  14.200 $\pm$ 1.933\\  
   DRIBO  &   &   &  9.883 $\pm$ 1.430 &  9.848 $\pm$ 1.493 &  4.910 $\pm$ 0.678 &  5.497 $\pm$ 0.874 &  5.931 $\pm$ 1.612 &  5.504 $\pm$ 1.110 &  12.091 $\pm$ 1.532\\  
     DS1  &   &   &   &  3.147 $\pm$ 1.209 &  9.673 $\pm$ 1.901 &  9.987 $\pm$ 1.878 &  6.762 $\pm$ 1.771 &  8.455 $\pm$ 1.494 &  9.150 $\pm$ 0.480\\  
     DS2  &   &   &   &   &  9.614 $\pm$ 1.753 &  9.902 $\pm$ 1.365 &  6.656 $\pm$ 1.682 &  8.362 $\pm$ 1.408 &  9.148 $\pm$ 0.423\\  
   DNET1  &   &   &   &   &   &  3.968 $\pm$ 0.513 &  3.929 $\pm$ 0.635 &  4.361 $\pm$ 1.307 &  12.254 $\pm$ 1.293\\  
   DNET2  &   &   &   &   &   &   &  5.982 $\pm$ 1.165 &  5.379 $\pm$ 0.963 &  12.728 $\pm$ 1.127\\  
   DNET3  &   &   &   &   &   &   &   &  3.953 $\pm$ 0.888 &  9.938 $\pm$ 1.085\\  
   DNET4  &   &   &   &   &   &   &   &   &  11.050 $\pm$ 1.109\\  
\et
\end{adjustwidth}
\etab

\newpage

\btab[!ht]
\tiny
\begin{adjustwidth}{-2.25in}{0in}
\caption{\bf Comparison of Different Gene-based and Network dendrograms for the second ensemble (10 sets of 30 organisms in each set).} \label{tab3}
\bt{c|ccccccccc}
DENDROGRAMS &  D1ENZ &  DRIBO &  DS1 &  DS2 &  DNET1 &  DNET2 &  DNET3 &  DNET4 &  DRAND\\ 
\hline
\hline
   DENZS  &  4.482 $\pm$ 1.040 &  6.635 $\pm$ 1.524 &  18.157 $\pm$ 0.709 &  18.202 $\pm$ 0.747 &  7.925 $\pm$ 2.445 &  8.458 $\pm$ 0.700 &  12.380 $\pm$ 1.701 &  10.129 $\pm$ 1.827 &  22.456 $\pm$ 0.835\\  
   D1ENZ  &   &  6.625 $\pm$ 1.752 &  17.028 $\pm$ 2.314 &  17.028 $\pm$ 2.379 &  8.333 $\pm$ 2.739 &  8.735 $\pm$ 0.874 &  11.728 $\pm$ 2.983 &  9.750 $\pm$ 2.659 &  21.295 $\pm$ 2.334\\  
   DRIBO  &   &   &  14.118 $\pm$ 1.629 &  14.140 $\pm$ 1.641 &  7.532 $\pm$ 1.480 &  8.392 $\pm$ 1.509 &  8.859 $\pm$ 2.307 &  7.123 $\pm$ 1.825 &  18.479 $\pm$ 1.452\\  
     DS1  &   &   &   &  5.046 $\pm$ 1.076 &  13.342 $\pm$ 2.805 &  15.033 $\pm$ 1.702 &  8.645 $\pm$ 1.449 &  11.584 $\pm$ 1.408 &  12.970 $\pm$ 0.897\\  
     DS2  &   &   &   &   &  13.328 $\pm$ 2.625 &  15.078 $\pm$ 1.717 &  8.700 $\pm$ 1.317 &  11.622 $\pm$ 1.253 &  13.028 $\pm$ 0.917\\  
   DNET1  &   &   &   &   &   &  6.347 $\pm$ 2.633 &  6.035 $\pm$ 1.762 &  5.976 $\pm$ 1.099 &  18.153 $\pm$ 2.231\\  
   DNET2  &   &   &   &   &   &   &  9.536 $\pm$ 2.679 &  8.036 $\pm$ 2.136 &  19.841 $\pm$ 1.785\\  
   DNET3  &   &   &   &   &   &   &   &  5.391 $\pm$ 1.401 &  14.121 $\pm$ 0.940\\  
   DNET4  &   &   &   &   &   &   &   &   &  16.337 $\pm$ 0.933\\  
\et
\end{adjustwidth}
\etab

\newpage


\section*{Additional Files}

\subsection*{ensembles.txt}

The file \texttt{ensembles.txt} contains the organisms in each set in each ensemble used in this work. The organisms are referred to by their KEGG code.

\subsection*{trees.txt}

The file \texttt{trees.txt} contains all dendrograms generated for each set in each ensemble. The dendrograms are given in \texttt{newick} format.

\end{document}